# UNET-TTS: IMPROVING UNSEEN SPEAKER AND STYLE TRANSFER IN ONE-SHOT VOICE CLONING

*Rui Li, Dong Pu, Minnie Huang, Bill Huang*

CloudMinds Inc., China
ryan.li@cloudminds.com

## ABSTRACT

One-shot voice cloning aims to transform speaker voice and speaking style in speech synthesized from a text-to-speech (TTS) system, where only a shot recording from the target reference speech can be used. Out-of-domain transfer is still a challenging task, and one important aspect that impacts the accuracy and similarity of synthetic speech is the conditional representations carrying speaker or style cues extracted from the limited references. In this paper, we present a novel one-shot voice cloning algorithm called Unet-TTS that has good generalization ability for unseen speakers and styles. Based on a skip-connected U-net structure, the new model can efficiently discover speaker-level and utterance-level spectral feature details from the reference audio, enabling accurate inference of complex acoustic characteristics as well as imitation of speaking styles into the synthetic speech. According to both subjective and objective evaluations of similarity, the new model outperforms both speaker embedding and unsupervised style modeling (GST) approaches on an unseen emotional corpus.

## 1. INTRODUCTION

Based on deep learning's modeling ability, text-to-speech (TTS) voice quality and naturalness have been greatly improved. Using multi-speaker and multi-style corpora, TTS is able to simulate expressive speech to support diverse customers, which enables various applications such as voice assistants, audiobooks, and voice-based customer service [1,2]. Transfer-learning allows a new TTS model to generate different speakers or styles with only a small amount of training data. Adapting a multi-speaker average model to a few text-audio pairs has been reported to achieve the same level as a single-speaker TTS [3-5]. Moreover, fine-tuning only selected parameters of the average model does not degrade speech quality, thereby reducing memory usage for each target speaker [3,6]. In addition, one-shot voice cloning, which neither updates any model parameters nor requires additional resources, has received a lot of interest [1]. A fixed-dimensional embedding vector from a speaker verification (SV) task is utilized as the speaker control condition when training the multi-speaker TTS model. In this case, a novel similar speaker voice can be created by sampling from the embedding prior with a target reference speech [3,7-9]. On the other hand, through the use of unsupervised style modeling technologies such as GST and VAE, style tokens or style latent variables are learned from the training data, and the style of a reference speech can be predicted [10-12]. It is possible to transform speech styles under these latent factors and control their intensity at run-time. In short, the additional conditional representations extracted from the limited references and subsequently employed in the model inference are crucial for determining the similarity and quality of synthetic speech [9,13].

Nevertheless, speaker embedding or unsupervised modeling have their own drawbacks as a result of limited and imbalanced training datasets. The speaker embedding in SV tends to preserve stable coarse-grained acoustic characteristics and excludes highly dynamic fine-grained features or random variations [13]. Accordingly, this method can only change speaker identity and rarely alter the style of speaking, such as speech rate, emotion, intonation, pause and stress. Also, the synthetic speech from an unseen speaker generally bears less resemblance to the original voice [7-9]. The method of unsupervised style modeling faces the problem that the number of captured styles is limited and their types cannot be controlled. Models sometimes analyze features that are irrelevant to speaking behavior, like noise and reverberation, and the inductive biases in unsupervised learning may instead cause confusion [10,14]. We discovered, for example, that models trained on corpora with a significant gender imbalance often alter the gender of synthetic target speech to the opposite. Likewise, this method cannot transfer the style unseen during training. Therefore, generating new speaker voices with unknown and arbitrary styles in one-shot TTS remains a challenge [1].

Unet-TTS is motivated by the desire to synthesize expressive speech in out-of-domain speaker identity and speaking style through one reference audio. This study shows that the new algorithm is able to effectively transfer speaker-level voice and emotion attributes, as well as utterance-level speaking styles, into synthetic speech. The proposed system consists of four components: *Duration Predictor* and *Content Encoder* produce phoneme duration and phoneme-level context hidden sequence respectively; *Style Encoder* is used to extract multi-level speaker and style representations from its different network layers; *Mel Decoder*, which forms a U-net structure with Style Encoder, reuses these multi-level latent vectors at its corresponding level layers and generates a spectrogram based on the content hiddens. The application of U-net in the image field has demonstrated its advantages in recovering information details and producing high resolution [15,16]. This has led it to show great potential in the field of speech processing as well, such as speech enhancement [17-20], audio separation [21-23] and voice conversion [24,25]. Compared to the previously reported speaker embedding or latent style factors, the multi-level representations of Unet-TTS provide much more detailed conditional information to facilitate inferences about complex features of speaker voice and speaking style. The synthetic examples[1] and code[2] are available online.

---

[1] https://cmsmartvoice.github.io/Unet-TTS
[2] https://github.com/CMsmartvoice/One-Shot-Voice-Cloning

The main contributions of this paper are summarized as follows:

1. A simple yet effective method is proposed for synthesizing a speech rate that matches the reference audio by adjusting the output of Duration Predictor according to the reference's duration statistics.

2. A content loss after Style Encoder is introduced to optimize the model jointly with the reconstruction loss, thereby enhancing the disentanglement between content and speaker-or-style.

3. The proposed model, based on the powerful spectrogram reconstruction of U-net network, is able to predict more precisely the details of acoustic features regarding pitch, harmonic, spectral envelope, and intensity. Some utterance-level speaking styles, including pause and stress, can be imitated into speech.

## 2. PROPOSED APPROACH

Our proposed Unet-TTS, illustrated in Fig.1(b), adopts the same architecture of speech duration independent modeling as that of traditional TTS or FastSpeech [26]. Phonemes are used as content inputs in the model, and their durations are modeled and predicted by the Duration Predictor. Content Encoder converts a one-hot-formed phoneme combination into a dense vector sequence with context, and then expands it into a longer one by repeating each vector depending on its duration. In parallel, Style Encoder extracts speaker and style embeddings from a reference spectrogram. Conditioning on these embeddings, Mel Decoder finally converts the expanded phoneme hiddens into a mel spectrogram. In addition, a pre-trained Content Encoder is selected and then frozen during training to supervise the last output of the Style Encoder, preventing the content information from leaking into the extracted embeddings. The pre-trained Content Encoder in this paper is obtained from a multi-speaker FastSpeech model using the speaker embedding approach, as shown in Fig.1(a). The speaker embeddings are derived in advance from a pre-trained speaker recognition model. According to [6], they are only employed in each Conditional Normalization module of the mel decoder, while Instance Normalization is used here.

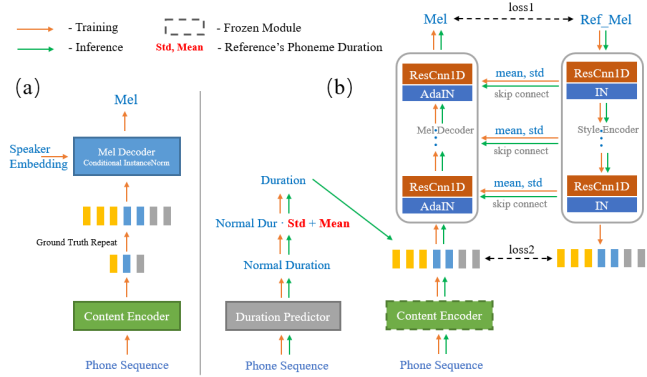

**Fig. 1.** *The structure of the pre-training model for Content Encoder (a), and Unet-TTS (b).*

### 2.1. Duration Predictor

The subjective perception of style similarity is heavily influenced by speech rate. In order to improve the transferability of speech rate in one-shot voice cloning, we use phoneme-level duration mean and standard deviation of the reference speech to adjust the output of Duration Predictor at run-time. Since the reference speech is the prediction target during training, it is equivalent to learning a speaker-independent phoneme to normalized duration mapping. The large-scale multi-speaker datasets ensure the model covers diverse text, and the normalized duration output reflects the intrinsic relative variation in length among local phonemes in certain speech segments.

### 2.2. Mel Decoder & Style Encoder

Mel Decoder and Style Encoder each contain an equal number of sub-modules, which together form a U-net structure. These sub-modules share the same structure, where ResCnn1D and Instance Normalization (IN) are the major components. But the forward of these modules is opposite on both sides of U-net. In the Style Encoder, a sequence of latent variables, i.e., the channel-wise mean and standard deviation of the hidden inputs in IN, can be successively extracted from different levels of the network. Then, they are reused to bias and scale the normalized hiddens along the Mel Decoder in its corresponding level IN layer, which is called adaptive instance normalization (AdaIN) according to [25,27]. Finally, Mel Decoder generates a mel spectrogram based on any content representation converted by the Content Encoder. The above process can be expressed as:

$$content = \text{Content\_Encoder}(phone\_seq)$$

$$\begin{cases} [means, stds] \\ content\_pred \end{cases} = \text{Style\_Encoder}(ref\_mel)$$

$$mel\_pred = \text{Mel\_Decoder}(content, [means, stds])$$

where ***content*** and ***content_pred*** are the outputs of the pre-trained Content Encoder and Style Encoder, respectively; ***mel_pred*** and ***ref_mel*** are the mel spectrogram generated by Mel Decoder and the reference spectrogram, respectively; ***means*** and ***stds*** are the extracted speaker and style embeddings from Style Encoder.

During training, ***ref_mel*** is the ground truth of ***mel_pred***, and ***content*** and ***content_pred*** are aligned in time. In addition to the spectrogram reconstruction L1 loss, there is a content disentangling loss computing the L2 error between ***content*** and ***content_pred***:

$$Loss = \|mel - mel\_pred\|_1 + \|content - content\_pred\|_2$$

## 3. EXPERIMENTS

In our experiments, a multi-speaker Mandarin speech corpora of 85 hours with 218 emotion-neutral speakers is used [28]. We split the dataset for training and testing: 188 speakers are used for training, 10 speakers for validation, and 20 speakers for cloning. Also, we select 10 Mandarin speakers from the Emotional Speech Dataset (ESD) [29], each with 5 emotion states (neutral, happy, angry, sad and surprise), for unseen voice and style transfer test. We use Montreal Forced Aligner [30] trained on these two datasets to align the speech with the text to get the duration of each phoneme. We use a sampling rate of 16 kHz and log-mel spectrograms with 80 bins, and apply the STFT with a FFT size of 800, hop size of 200, and window size of 800 samples.

The pre-trained model for Content Encoder is similar in architecture to AdaSpeech [6], and Instance Normalization is used here. Duration Predictor consists of one-layer self-attention network and two-layer 1D convolutional network with ReLU activation, and an extra linear layer to output a scalar. The Mel Decoder and Style Encoder have six successive sub-modules each.

The ResCnn1D block adds its input to the output of its internal Cnn1D module, and the Cnn1D module consists of two-layer 1D convolutional network with a ReLU activation between them. The dimension of phoneme hidden variables, the hidden size of the self-attention, and the filters of all the 1D convolutions are set to 256. The kernel sizes of the 1D convolution of Content Encoder and Duration Predictor are set to 3, while those of Mel Decoder and Style Encoder are set to 9.

To compare the performance, we trained two one-shot TTS baseline models based on the Tacotron architecture [31] using the same training corpora. They use the speaker embedding in SV and the style embedding from an unsupervised style modeling named GST [10] to provide speaker and style information, respectively. In the following section, they are identified by **Tacotron_spkembed** and **Tacotron_GST**, respectively.

## 4. RESULTS AND EVALUATION

### 4.1. Reconstruction

In TTS, text-to-spectrogram conversion is a one-to-many mapping problem, and the model is mainly optimized by minimizing a spectrogram reconstruction loss. During one-shot voice cloning, the extracted low-dimensional speaker or style representation is one of the factors responsible for decreasing the loss in such a cross-domain conversion, which contains critical spectrogram information and is used by the model to reconstruct the spectrogram from text. An incomplete conditional encoding will lead to a serious degradation in reconstruction accuracy. Due to operations such as down sampling and pooling, speaker embedding or style tokens that are derived from one layer of the network are likely to lose some details that the model cannot reconstruct. In contrast, the U-net network has the advantage of incorporating hierarchical representations of the encoder network into the reconstruction process, which prevents the missing of potentially critical features. As can be seen from Fig.2(a), the U-net based one-shot TTS achieves a significant improvement in reducing reconstruction loss. The GST algorithm, which also uses all the information of the spectrogram as a reference, shows only a small drop in loss compared to the speaker embedding method.

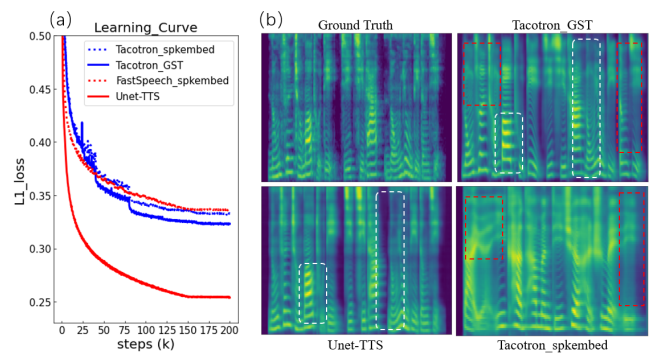

**Fig. 2.** *The reconstruction loss vs. steps of the validation set (a) and the generated log-mel spectrograms of the same text (b) by different models.*

Speaking behavior is highly variable because of both voluntary and involuntary activity. From a spectral perspective, the acoustic characteristics about speaker voice and speaking style can be roughly grouped into three categories [2,6]. The first are the intrinsic and stable coarse-grained acoustic features, which are entirely determined by the speaker's physiological structure and can be used to identify the speaker. They represent the low-frequency components of a spectrogram, including the average pitch, the spectral envelope reflecting vocal-tract response, and the relative amplitude and position of formants. Their individual differences will directly impact the subjective perception of speaker identity similarity. Second, spectrograms can reveal acoustically unstable characteristics, such as sharp and rapid variations in duration, pitch, intensity, and spectral fine-structure, all of which are susceptible to the speaker's psychological changes. Speakers can convey different emotions or intentions by altering these features locally. They influence the subjective perception of both speaker similarity and style similarity. Lastly, spectrograms show some speaking styles that are not related to the speaker identity, including pause, and stress, etc. Appropriate filling of these elements into speech can improve the speech's naturalness as well as the subjective perception of style similarity. Fig.2(b) shows the generated spectrograms of different models based on a reference audio of the validation, where the text to synthesize matches the text of the reference. Since the speaker embedding can hardly provide fine-grained spectral information, its mel spectrogram is rather smooth and lacks many details. Large flat areas are present, especially in the mid- and high-frequency bands (marked by the red dashed line). GST employs the entire spectrogram as a style input, resulting in a higher resolution spectrogram output with well-defined edges and contours. Nevertheless, some phonemes have inaccurate pitch and intensity predictions, and no apparent pause boundary can be recognized (marked by white dashed line). As expected, Unet-TTS is much more accurate in predicting the fine-grained acoustic features, and can even mimic the utterance-level pause style.

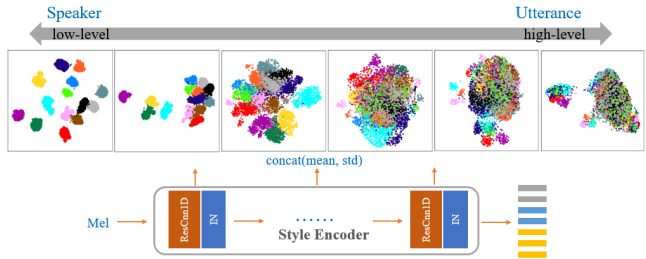

**Fig. 3.** *The visualization of latent embedding extracted from each IN layer of different levels in Style Encoder.*

We also concatenate the mean and standard deviation vectors for each IN layer in Style Encoder and visualize these deep features with principal component analysis in a two-dimensional plane. Fig.3 shows the arbitrary text results for 14 randomly selected unseen speakers, each with 200 samples. Despite the fact that speaker identifier labels are not used in the training process, Style Encoder's low-level network automatically learns a meaningful embedding space with clear speaker identity groups. But as the network deepens, these latent distributions gradually become indistinguishable under speaker identifier labels. At the highest level, the distributions of all speakers overlapped almost identically. In our view, this shift process may be explained by the three types of acoustic spectrogram features discussed above: the low-level networks extract stable coarse-grained characteristics related to the speaker, while the high-level networks capture unstable fine-grained ones specific to the current utterance. This is a hypothesis that has yet to be confirmed. If this is the case, it would be possible to combine different levels of the attributes about speaker and style from several utterances into the same Mel Decoder, allowing Unet-TTS to synthesize a speech with the voice of one speaker but the style of another. Our future research will focus on this issue.

### 4.2. Voice and Style Transfer

Large corpora of collected multi-speaker recordings often lack richness of emotion and are at a neutral level in speed, pitch and intensity. For TTS models trained on such data, it becomes difficult to synthesize arbitrary text in an unseen emotion. Thus, it motivates us to investigate the transfer effectiveness of Unet-TTS for out-of-domain speaker and style. Fig.4 presents the distributions of phoneme duration, phoneme-level average energy and average pitch (F0), as well as the mel-cepstral distortion (MCD) [32] scores of the synthetic speech by different models. Five emotions ("angry," "happy," "neutral," "sad," and "surprise") are tested, and 200 samples are randomly selected for each speaker with each emotion. All spectrograms are converted to waveforms using a Multi-band MelGAN vocoder [33]. F0 and spectral envelope cepstrum are extracted using the WORLD vocoder [34]. The MCD in this paper is calculated by first averaging the cepstrum along the time axis and then comparing the error against the ground truth, which can be used to evaluate the voice similarity.

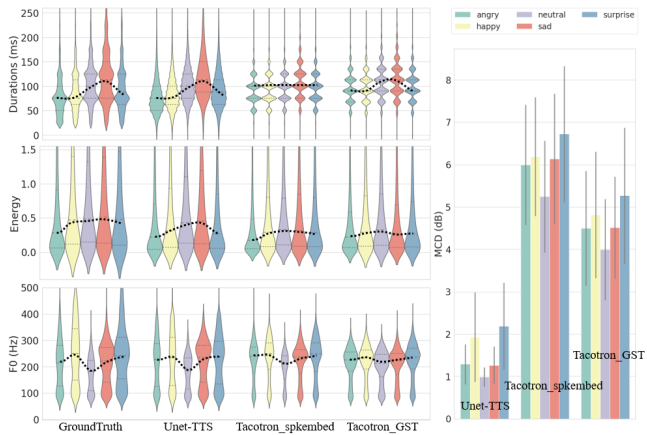

**Fig. 4.** *The distribution of phoneme-level energy, duration, pitch, as well as the MCD of synthetic speech.*

The proposed Duration Predictor, which uses the reference speech's duration statistics to adjust its output, produces distributions that are close to those of the reference (ground-truth). But the duration distributions derived from Tacotron's attention scores are narrower and even show only a few isolated peaks. This suggests that Tacotron's synthetic speech rates seem to be restricted to a few fixed ranges, which may negatively impact the rhythm similarity. Again, the phoneme-energy and –F0 distributions of Unet-TTS are both wider than those of speaker embedding or GST techniques. It appears Unet-TTS is more effective in transferring these acoustic features of the out-of-domain emotional speech, despite the absence of these speakers or styles in the training data. Moreover, Unet-TTS achieves the lowest MCD score, indicating that its synthetic speech is closest to the reference in terms of voice similarity, and its spectral envelope is more representative of the speaker's vocal-tract response.

In addition, the figure shows the median values of duration, energy, and pitch distributions in relation to emotion type (marked by black dashed line). Because the speaker embedding provides little information about speech rate, its duration output does not vary between emotions, suggesting that this model is ineffective in transferring speech rate. In exception to the above, the variations in duration, phoneme-energy, and phoneme-F0 among emotions of each model more or less follow the trends found in the reference speech. Nevertheless, the degree of variation by U-net TTS is closer to that of the ground truth speech, whereas neither speaker embedding nor GST changes significantly. Unet-TTS is therefore more capable of capturing and generating highly variable acoustic features, independent of the training data.

**Table 1.** *Speech naturalness MOS with 95% confidence intervals.*

| Method | GT | GT Vocoder | Tacotron spkembed | Tacotron GST | Unet-TTS |
|---|---|---|---|---|---|
| MOS | 3.99 | 3.85 | 3.34 | 3.67 | 3.62 |

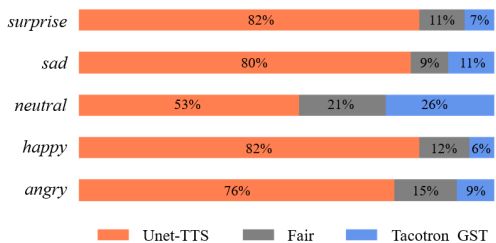

**Fig. 5.** *Pairwise similarity preference test for emotion transfer.*

Next, we perform two subjective evaluations on the above synthetic speech: naturalness MOS and pairwise similarity preference. A total of 20 Mandarin-speaking listeners were invited, and each listener was presented with 20 samples or pairs from each source. In the MOS test, the naturalness is rated on a scale of 1–5 in intervals of 0.5. In the pairwise similarity preference test, a reference speech and two speech generated from Unet-TTS and GST were given, and the participants were asked to choose the one closer to both speaker and style of the reference audio. As shown in Table 1, the MOS score of Unet-TTS is slightly lower than that of GST, and both are significantly higher than the speaker embedding method. We found that in Unet-TTS, short reference audio may cause blurred or incorrect pronunciation, and poor-quality audio may produce some noise or mechanical sounds, which are the causes of MOS degradation. In contrast, as shown in Fig.5, Unet-TTS completely outperforms GST-Tacotron in terms of emotional similarity, which we believe is remarkable. In most cases, GST fails to transfer speaker identity or speaking style from the target speech, except for neutral emotion. These observations validate the effectiveness of our proposed model in imitating and generating the out-of-domain emotions.

### 5. CONCLUSIONS

In this work, we propose a U-net based one-shot TTS algorithm with powerful transferability for unseen speaker voices and speaking styles. The new model outperforms the baseline models in both subjective and objective evaluations, especially when it comes to imitating unseen emotions. The results of this study indicate that Unet-TTS has great potential for tackling the problem of one-shot out-of-domain transfer. In future work, we will search for the meaning of extracted multi-level speaker and style embeddings, and investigate the possibility of controllable multi-utterance style transfer TTS.

## 6. REFERENCES

[1] Q. Xie, X. Tian, G. Liu *et al.*, "The multi-speaker multi-style voice cloning challenge 2021," in *IEEE International Conference on Acoustics, Speech and Signal Processing*, 2021: IEEE, pp. 8613-8617.

[2] X. Tan, T. Qin, F. Soong *et al.*, "A survey on neural speech synthesis," *arXiv preprint arXiv:.15561,* 2021.

[3] S. O. Arik, J. Chen, K. Peng *et al.*, "Neural voice cloning with a few samples," in *NeuralPS*, 2018, vol. 31, pp. 10040-10050.

[4] Y. Deng, L. He, and F. Soong, "Modeling multi-speaker latent space to improve neural tts: Quick enrolling new speaker and enhancing premium voice," *arXiv preprint arXiv:.05253,* 2018.

[5] H. B. Moss, V. Aggarwal, N. Prateek *et al.*, "Boffin tts: Few-shot speaker adaptation by bayesian optimization," in *IEEE International Conference on Acoustics, Speech and Signal Processing*, 2020: IEEE, pp. 7639-7643.

[6] M. Chen, X. Tan, B. Li *et al.*, "Adaspeech: Adaptive text to speech for custom voice," in *International Conference on Learning Representations*, 2021.

[7] Y. Jia, Y. Zhang, R. J. Weiss *et al.*, "Transfer learning from speaker verification to multispeaker text-to-speech synthesis," in *32nd International Conference on Neural Information Processing Systems*, 2018, pp. 4485-4495.

[8] Z. Cai, C. Zhang, and M. Li, "From speaker verification to multispeaker speech synthesis, deep transfer with feedback constraint," in *Interspeech*, 2020, pp. 3974-3978.

[9] E. Cooper, C.-I. Lai, Y. Yasuda *et al.*, "Zero-shot multi-speaker text-to-speech with state-of-the-art neural speaker embeddings," in *IEEE International Conference on Acoustics, Speech and Signal Processing*, 2020: IEEE, pp. 6184-6188.

[10] Y. Wang, D. Stanton, Y. Zhang *et al.*, "Style tokens: Unsupervised style modeling, control and transfer in end-to-end speech synthesis," in *International Conference on Machine Learning*, 2018: PMLR, pp. 5180-5189.

[11] Y.-J. Zhang, S. Pan, L. He *et al.*, "Learning latent representations for style control and transfer in end-to-end speech synthesis," in *IEEE International Conference on Acoustics, Speech and Signal Processing*, 2019: IEEE, pp. 6945-6949.

[12] G. Sun, Y. Zhang, R. J. Weiss *et al.*, "Generating diverse and natural text-to-speech samples using a quantized fine-grained vae and autoregressive prosody prior," in *IEEE International Conference on Acoustics, Speech and Signal Processing*, 2020: IEEE, pp. 6699-6703.

[13] C.-M. Chien, J.-H. Lin, C.-y. Huang *et al.*, "Investigating on incorporating pretrained and learnable speaker representations for multi-speaker multi-style text-to-speech," in *IEEE International Conference on Acoustics, Speech and Signal Processing*, 2021: IEEE, pp. 8588-8592.

[14] F. Locatello, S. Bauer, M. Lucic *et al.*, "Challenging common assumptions in the unsupervised learning of disentangled representations," in *International Conference on Machine Learning*, 2018: PMLR, pp. 4114-4124.

[15] O. Ronneberger, P. Fischer, and T. Brox, "U-net: Convolutional networks for biomedical image segmentation," in *International Conference on Medical image computing and computer-assisted intervention*, 2015: Springer, pp. 234-241.

[16] X. Hu, M. A. Naiel, A. Wong *et al.*, "RUNet: A robust UNet architecture for image super-resolution," in *IEEE/CVF Conference on Computer Vision and Pattern Recognition Workshops*, 2019, pp. 0-0.

[17] C. Macartney and T. Weyde, "Improved speech enhancement with the wave-u-net," *arXiv preprint arXiv:.11307,* 2018.

[18] H.-S. Choi, J.-H. Kim, J. Huh *et al.*, "Phase-aware speech enhancement with deep complex u-net," in *International Conference on Learning Representations*, 2018.

[19] V. Kothapally, W. Xia, S. Ghorbani *et al.*, "Skipconvnet: Skip convolutional neural network for speech dereverberation using optimally smoothed spectral mapping," in *Interspeech*, 2020, pp. 2020-2048.

[20] J.-H. Kim and J.-H. Chang, "Attention Wave-U-Net for Acoustic Echo Cancellation," in *Interspeech*, 2020, pp. 3969-3973.

[21] A. Jansson, E. Humphrey, N. Montecchio *et al.*, "Singing voice separation with deep u-net convolutional networks," in *18th International Society for Music Information Retrieval Conference*, 2017.

[22] D. Stoller, S. Ewert, and S. Dixon, "Wave-u-net: A multi-scale neural network for end-to-end audio source separation," in *19th International Society for Music Information Retrieval Conference*, 2018, pp. 334-340.

[23] G. Meseguer-Brocal and G. Peeters, "Conditioned-U-Net: Introducing a control mechanism in the U-Net for multiple source separations," in *20th International Society for Music Information Retrieval Conference*, 2019.

[24] D.-Y. Wu, Y.-H. Chen, and H.-Y. Lee, "VQVC+: One-Shot Voice Conversion by Vector Quantization and U-Net architecture," in *Interspeech*, 2020, pp. 4691-4695.

[25] Y.-H. Chen, D.-Y. Wu, T.-H. Wu *et al.*, "Again-vc: A one-shot voice conversion using activation guidance and adaptive instance normalization," in *IEEE International Conference on Acoustics, Speech and Signal Processing*, 2021: IEEE, pp. 5954-5958.

[26] Y. Ren, Y. Ruan, X. Tan *et al.*, "Fastspeech: Fast, robust and controllable text to speech," in *33rd International Conference on Neural Information Processing Systems*, 2019, p. 10.

[27] X. Huang and S. Belongie, "Arbitrary style transfer in real-time with adaptive instance normalization," in *IEEE International Conference on Computer Vision*, 2017, pp. 1501-1510.

[28] Y. Shi, H. Bu, X. Xu *et al.*, "Aishell-3: A multi-speaker mandarin tts corpus and the baselines," in *Interspeech*, 2021.

[29] K. Zhou, B. Sisman, R. Liu *et al.*, "Seen and unseen emotional style transfer for voice conversion with a new emotional speech dataset," in *IEEE International Conference on Acoustics, Speech and Signal Processing*, 2021: IEEE, pp. 920-924.

[30] M. McAuliffe, M. Socolof, S. Mihuc *et al.*, "Montreal Forced Aligner: Trainable Text-Speech Alignment Using Kaldi," in *Interspeech*, 2017, vol. 2017, pp. 498-502.

[31] J. Shen, R. Pang, R. J. Weiss *et al.*, "Natural tts synthesis by conditioning wavenet on mel spectrogram predictions," in *IEEE International Conference on Acoustics, Speech and Signal Processing*, 2018: IEEE, pp. 4779-4783.

[32] R. Kubichek, "Mel-cepstral distance measure for objective speech quality assessment," in *IEEE Pacific Rim Conference on Communications Computers and Signal Processing*, 1993, vol. 1: IEEE, pp. 125-128.

[33] G. Yang, S. Yang, K. Liu *et al.*, "Multi-band MelGAN: Faster waveform generation for high-quality text-to-speech," in *IEEE Spoken Language Technology Workshop (SLT)*, 2021: IEEE, pp. 492-498.

[34] M. Morise, F. Yokomori, and K. Ozawa, "WORLD: a vocoder-based high-quality speech synthesis system for real-time applications," *IEICE TRANSACTIONS on Information and Systems,* vol. 99, no. 7, pp. 1877-1884, 2016.